\documentstyle[11pt,aaspp4]{article}
\def\gsim{\;\rlap{\lower 2.5pt
 \hbox{$\sim$}}\raise 1.5pt\hbox{$>$}\;}
\def\lsim{\;\rlap{\lower 2.5pt
   \hbox{$\sim$}}\raise 1.5pt\hbox{$<$}\;}
\newcommand\beq{\begin{equation}}
\newcommand\eeq{\end{equation}}
\def\lya{Ly$\alpha$~}

\begin{document}

\title{Direct Measurement of Cosmological Parameters from the Cosmic
Deceleration of Extragalactic Objects}
\author{Abraham Loeb}
\medskip
\affil{Harvard-Smithsonian Center for Astrophysics, 60 Garden Street,
Cambridge, MA 02138}

\begin{abstract}

The redshift of all cosmological sources drifts by a systematic velocity of
order a few ${\rm m~s^{-1}}$ over a century due to the deceleration of the
Universe. 
The specific functional dependence of the predicted velocity shift on the
source redshift can be used to verify its cosmic origin, and to measure
directly the values of cosmological parameters, such as the density
parameters of matter and vacuum, $\Omega_{\rm M}$ and $\Omega_{\Lambda}$,
and the Hubble constant $H_0$.  For example, an existing spectroscopic
technique, which was recently employed in planet searches, is capable of
uncovering velocity shifts of this magnitude. The cosmic deceleration
signal might be marginally detectable through two observations of $\sim
10^2$ quasars set a decade apart, with the HIRES instrument on the Keck 10
meter telescope.  The signal would appear as a global redshift change in
the Ly$\alpha$ forest templates imprinted on the quasar spectra by the
intergalactic medium.  The deceleration amplitude should be isotropic
across the sky.  Contamination of the cosmic signal by peculiar
accelerations or local effects is likely to be negligible.

\end{abstract}

\keywords{cosmology: theory}

\centerline{submitted to {\it ApJ Letters}, Feb. 10th, 1998}

\section{Introduction}

Most conventional methods for measuring the values of cosmological
parameters rely on the determination of the luminosity distance to
extragalactic sources with known redshifts. Consequently, these methods
must make {\it apriori} assumptions about the intrinsic luminosity of the
sources and are often plagued by evolutionary effects or systematic
uncertainties [such problems are particularly acute for galaxies (Tinsley
1977) but less so for Type-Ia supernovae (Garnavich et al. 1997; Perlmutter
et al. 1998)].  The alternative approach of using the spectrum of microwave
background anisotropies for the same purpose (Zaldarriaga, Spergel, \&
Seljak 1997; Bond, Efstathiou \& Tegmark 1997, and references therein)
suffers from partial degeneracies between different cosmological parameters
and might be compromised by foreground source contamination.  In this {\it
Letter} I show that a direct measurement of the deceleration of the
Universe is not far out of reach of current technology and might be used to
determine the values of the cosmological parameters in the future.

The search for a small reflex motion of stars due to potential planetary
companions has recently led to the development of an advanced spectroscopic
technique which yields radial velocity errors of $\sim 3~{\rm m~s^{-1}}$
(Butler et al.  1996, and references therein). The technique utilizes a
fast echelle spectrograph and an iodine absorption cell placed at the
spectrometer entrance slit. The superimposed iodine lines provide a
reference wavelength frame against which the radial velocity shifts of the
stars are measured.  The technique currently reaches a sensitivity of $\sim
3~{\rm m~s^{-1}}$ for a 10 min exposure on a $B\sim 6$ mag star using a 3
meter telescope and is primarily limited by photon statistics.
Extrapolation of this sensitivity to 100 hours of integration on the Keck
10 meter telescope should allow for the detection of a similar velocity
shift
in the spectrum of a $B\sim 16$ mag source.  In \S 2 I will show that a
velocity shift of a few ${\rm m~s^{-1}}$ is expected to occur over a
century in the spectrum of extragalactic objects at high redshifts, merely
due to the deceleration of the Universe.  Measurement of this shift for a
sufficiently large statistical sample of extragalactic objects could then
be used to determine the values of the fundamental parameters which define
the geometry of the Universe. As an illustrative example, I will consider
in \S 3 the expected redshift drift of the \lya forest in quasar spectra.
Contamination of the cosmic deceleration signal by local peculiar
accelerations will be shown to be negligible.

\section{The Cosmic Signal}

We would like to evaluate the expected variation of the cosmic redshift of
a particular extragalactic source with time. For simplicity, let us assume
at first that the source does not possess any peculiar velocity or peculiar
acceleration, so that it maintains a fixed comoving coordinate ($dr_{\rm
s}=0$).  The radiation emitted by the source at two different times $t_{\rm
s}$ and $t_{\rm s}+\Delta t_{\rm s}$ will be observed at later times
$t_{\rm o}$ and $t_{\rm o}+\Delta t_{\rm o}$, related by (Weinberg 1972)
\begin{equation}
\int_{t_{\rm s}}^{t_{\rm o}} {dt \over a(t)} = 
\int_{t_{\rm s}+\Delta t_{\rm s}}^{t_{\rm o}+\Delta t_{\rm o}} 
{dt \over a(t)}~,
\label{eq:one}
\end{equation}
where $a(t)=(1+z)^{-1}$ is the scale factor of the Universe, normalized to
unity at zero redshift when the age of the Universe is $t_{\rm o}$.  For
$(\Delta t/t)\ll 1$, equation~(\ref{eq:one}) yields the standard redshift
factor between the observation and emission time intervals $\Delta t_{\rm
s}= [a(t_{\rm s})/a(t_{\rm o})] \Delta t_{\rm o}$. The source redshift
changes during this time interval by the amount
\begin{equation}
\Delta z\equiv 
{a(t_{\rm o}+\Delta t_{\rm o})\over a(t_{\rm s}+\Delta t_{\rm s})}-
{a(t_{\rm o})\over a(t_{\rm s})}\approx 
\left[{{\dot a}(t_{\rm o})-{\dot a}(t_{\rm s})\over 
a(t_{\rm s})}\right]\Delta t_{\rm o}= 
\left({{\dot z}_{\rm s}\over (1+z_{\rm s})}- 
(1+z_{\rm s}){\dot z}_{\rm o}\right)
\Delta t_{\rm o} ,
\label{eq:two}
\end{equation}
where an overdot indicates time derivative, and the second equality was
derived by Taylor expanding the scale factor to leading order, $a(t+\Delta
t)\approx a(t)+{\dot a}(t)\Delta t$. Note that the redshift change, $\Delta
z\propto [{\dot a}(t_{\rm o})-{\dot a}(t_{\rm s})]$, results from the
evolution of ${\dot a}$ with cosmic time, namely from the deceleration of
the Universe.  The Friedmann equations can be used to relate ${\dot z}=
-(1+z)^2 {\dot a}$ to the matter content of the Universe, yielding ${\dot
z}=-H_0(1+z)\left[\Omega_{\rm M}(1+z)^3 +\Omega_{\rm R}(1+z)^2+
\Omega_\Lambda\right]^{1/2}$; where $H_0\equiv {\dot a}(t_{\rm o})/a(t_{\rm
o})$ is the Hubble constant, $\Omega_{\rm M}$ and $\Omega_{\Lambda}$ are
the cosmological density parameters of matter and vacuum (the cosmological
constant), respectively, and $\Omega_{\rm R}\equiv (1-\Omega_{\rm
M}-\Omega_\Lambda)$. In summary, we get a spectroscopic velocity shift
\begin{equation}
\Delta v\equiv {c\Delta z\over (1+z_{\rm s})}=-\left\{\left[
\Omega_{\rm M}(1+z_{\rm s}) + \Omega_{\rm R}
+\Omega_\Lambda(1+z_{\rm s})^{-2}\right]^{1/2}-1\right\} 
H_0\Delta t_{\rm o} c~,
\label{eq:three}
\end{equation}
where $c$ is the speed of light.  This shift vanishes for an empty Universe
($\Omega_{\rm M}=0$, $\Omega_{\Lambda}=0$), and turns positive for an
inflating $\Omega_{\Lambda}$--dominated Universe.  For
$\Omega_{\Lambda}=0$, we obtain $\Delta v = -2~{\rm m~s^{-1}}~h_{65}
[{\sqrt{1+\Omega_{\rm M}z_{\rm s}}}-1] (\Delta t_{\rm o}/10^2~{\rm yr})$;
where $h_{65}= (H_0/65~{\rm km~s^{-1}~Mpc^{-1}})$.  Examples of
extragalactic sources for which a detection of the cosmic signal might be
feasible will be discussed in \S 3.

Finally, we consider contaminating effects. Any change in the comoving
coordinate of the source due to a peculiar velocity, $\delta v$, would only
modify the redshift change in equation~(\ref{eq:three}) by a correction of
order $\sim (\delta v/c)\ll 1$ (assuming $z_{\rm s}\ga 1$).  This
correction is only a fraction of a percent, given the characteristic
amplitude of peculiar flows in the Universe.  The effect of peculiar
accelerations could be more significant. First, we note that the peculiar
acceleration of the Sun relative to the Galaxy is comparable to the cosmic
signal, but can be easily separated from it based on its known direction
and magnitude (as the cosmic deceleration must be isotropic).
The characteristic amplitude of the remaining cosmic deceleration signal
is $\sim H_0 c$. In comparison, the typical acceleration inside a
virialized object of a density $\delta$, in units of the critical density,
and a velocity dispersion $\sigma$ is only a fraction $\sim (\sigma/c)
{\sqrt{\delta }}$ of the cosmic signal.  Given the characteristic values of
$(\sigma/c)\sim 10^{-3}$--$10^{-2}$ in galaxies or galaxy clusters, it is
clear that the peculiar acceleration can compete with the cosmic
deceleration signal only when $\delta \ga 10^4$, i.e. in the very dense
cores of virialized objects.  Most of the mass in the Universe resides in
the outer envelopes of such objects, where the peculiar accelerations are
much smaller than the cosmic deceleration signal.  Moreover, as the
peculiar accelerations are induced by local effects and hence have a random
sign, their significance could be further diminished by averaging over a
sufficiently large statistical sample of extragalactic objects.

\section{Illustrative Examples}

There are probably several innovative ways to measure the velocity drift
predicted by equation~(\ref{eq:three}). Below I consider two
straightforward examples to illustrate the feasibility of such a
measurement.

Luminous quasars are the brightest extragalactic sources, but the large
width of their broad emission lines ($\ga 10^3~{\rm km~s^{-1}}$) and their
substantial intrinsic variability renders the use of their spectral lines
for the detection of the cosmic deceleration signal impractical.  However,
the rich absorption line forest imprinted on the continuum flux of quasars
by the intergalactic medium provides an ideal template for this purpose.
The width of the \lya absorption lines is only $\sim 20~{\rm km~{s^{-1}}}$,
and the metal lines are even narrower.  There are hundreds of detectable
\lya lines per unit redshift down to HI column densities $\sim 10^{13}~
{\rm cm^{-2}}$ (Tytler 1998).  The existence of absorption lines across the
entire range of redshifts out to the quasar allows one to separate the
cosmic deceleration signal from contaminating effects, based on the
particular functional dependence on redshift predicted by
equation~(\ref{eq:three}).  Moreover, the cosmic signal should be isotropic
across the sky in difference from any local effect.  In total, there are
$\sim 10^2$ quasars with $B\sim 16$ mag over the sky, for which the
\lya forest can be observed from the ground, i.e. with $z_{\rm s}\ga 2.5$
(Hartwick \& Schade 1990).  The cosmic signal can then be searched for
through a cross-correlation analysis of the \lya forest template in the
spectrum of these quasars, taken at two different times with the HIRES
instrument on the Keck 10 meter telescope. Because the velocity drift is a
function of redshift, the cross-correlation analysis should be done by
using the functional form predicted by equation~(\ref{eq:three}) as a
kernel, and by exploring different possible values of the cosmological
parameters.  To illustrate the feasibility of this project, we first
consider a time separation of a century, namely $\Delta t_{\rm o}=100~{\rm
yr}$.  Each pixel in the Keck HIRES spectrum is 2~${\rm km~s^{-1}}$ wide.
According to equation~(\ref{eq:three}) the cosmic velocity shift would be
$\sim 10^{-3}$ of a pixel width. Assuming that the
\lya forest flux varies by $\sim 10\%$ from pixel to pixel, the overall
signal is at the level of $\sim 10^{-4}$.  Hence, with a signal-to-noise
ratio of $\sim 100$ per pixel, one needs $\sim (100\times 10^{-4})^{-2}=
10^4$ pixels to detect the cosmic deceleration signal at 1$\sigma$. More
than this number of pixels is generally available per quasar, and so the
signal is potentially detectable over a century for a single quasar.
Ignoring systematic instrumental limitations, it should therefore be
feasible to detect the cosmic signal only over a decade ($\Delta t_{\rm
o}=10~{\rm yr}$) by extending the sample to include $\sim 10^2$ such
quasars. (The signal to noise ratio increases as the square root of the
number of uncorrelated quasars).  The low HI column-density absorbers ($\la
10^{13}~{\rm cm^{-2}}$) are believed to be associated with underdense
regions and hence their peculiar accelerations should be lower than the
cosmic deceleration $\sim H_0 c$ by at least the ratio of the
corresponding void size to the Hubble length.  Changes in the HI absorption
line template due to the evolution of structure in the intergalactic medium
would result in negligible peculiar accelerations [$\sim (\delta v/c)
{\sqrt{\delta}} H c$] except for the extremely overdense, and hence rare,
regions which yield damped \lya absorption.

Obviously, the efficiency of this technique can improve dramatically if a
subset of all absorption lines (e.g., some metal lines) possess
particularly sharp spectral features. Nevertheless, we should also make
several cautionary remarks. First, extraction of cosmological parameters
requires redshift binning which will reduce the signal-to-noise ratio per
bin.  Detection of the effect at several standard deviations would require
an improvement by at least an order of magnitude in sensitivity over
existing instruments. Systematic effects will ultimately limit the
performance of any advanced instrumentation.  In addition, the \lya forest
lines compose a somewhat inferior spectral template for the purpose of
measuring velocity drifts compared to stellar lines, because of their lower
abundance, their somewhat larger width, and the potential variability of
their continuum source.

The other distant sources of interest are galaxies.  The emission spectra
of galaxies are merely a superposition of their stellar constituents.
However, the rich emission and absorption line template which is employed
for the detection of the slight reflex motion of individual stars (Butler
et al.  1996)
is degraded when incorporated into the galactic spectrum, due to the
broadening of the stellar lines by the velocity dispersion or rotation of
the galaxy (typically of order hundreds of ${\rm km~s^{-1}}$) and the
resulting blending of some lines. On the other hand, the cumulative
spectrum of a galaxy is much more stable than that of an individual star,
because the variability due to convective or atmospheric motions that
limits the radial velocity precision for individual stars, is averaged out
over the large number of uncorrelated changes in the stars which make up a
galaxy. Galaxies are much more abundant, and hence compose a larger
statistical sample, than quasars.

Since galaxies are typically $\sim 2$ orders of magnitude fainter than
bright quasars, it is necessary to adopt time intervals $\Delta t_{\rm
o}\sim 10^{3}~{\rm years}$ in order to detect the cosmic signal in their
spectra.  While such time intervals might appear impractical on the scale
of a human lifetime, they are accessible through the multiple images of a
background galaxy that is gravitationally lensed by a foreground cluster
of galaxies. The characteristic time delay between the two images produced
by a singular isothermal sphere of a 1D velocity dispersion $\sigma$ is
(Schneider, Ehlers, \& Falco 1992),
\begin{equation}
\Delta t_{\rm o}= \left(4\pi {\sigma^2\over c^2}\right)^2 {D_{\rm l} D_{\rm ls}
\over c D_{\rm s}} \left( 1+z_{\rm l}\right) 2y ,
\end{equation}
where $D_{\rm l}$, $D_{\rm ls}$, and $D_{\rm s}$, are the angular diameter
distances between the observer and the lens, the lens and the source, and
the observer and the source, respectively; $z_{\rm l}$ is the lens
redshift, and $y$ is the unlensed source position in units of the critical
radius of the lens. For the characteristic velocity dispersion of rich
clusters $\sigma \sim (1000$--$1500)~{\rm km~s^{-1}}$, the delay is just in
the range of $\Delta t_{\rm o}\sim 10^2$--$10^3$ years required for
detection of the cosmic deceleration signal. Moreover, the magnification
of the background galaxy by the cluster would boost its observed flux and
enhance the sensitivity to potential shifts in its spectral features.
Comparison of the spectra of multiple images of the same galaxy which are
delayed relative to each other should show the redshift difference
predicted by equation~(\ref{eq:three}). Unfortunately, the practicality of
measuring this redshift difference is impeded by the need to collect all
the light from the extended galaxy arclets within the observational
aperture (this is made more difficult by the fact that lensing conserves
surface brightness and stretches the area of the background galaxy on the
sky); missing light from some parts of the galaxy could lead to a
systematic velocity offset which is far greater than the cosmic signal.  In
addition, the cosmic signal is likely to be swamped by the change in the
image redshifts due to a transverse peculiar velocity of the lens; this
change is of order the product of the deflection angle and the transverse
velocity of the cluster, i.e. $\sim 10^{-4}\times 10^{-3} c= 30~{\rm
m~s^{-1}}$ (Pen 1998).

Note that lensing cannot be employed to search for velocity shifts in the
\lya forest because the different images of a lensed quasar follow different 
spatial paths through the intergalactic medium on their way to the
observer.

Finally, I should mention the somewhat speculative possibility of using
radio sources which offer exceptional frequency stability over a narrow
band width.  Examples for such sources include powerful masers in galactic
nuclei (Miyoshi et al. 1995), young extragalactic pulsars, or 21 cm
emission from cold gas at high redshifts.  Some of these sources (e.g.,
maser disks or pulsars in binary systems) might possess high peculiar
accelerations intrinsically; however, the feasibility of detecting a signal
with an amplitude as small as the cosmic deceleration from pulsar timing
was already demonstrated in the Milky Way galaxy (Damour \& Taylor 1991;
Stairs et al.  1997). The faintness of these sources might restrict such
measurements to the local Universe, but still provide a measurable signal
-- especially for an $\Omega_{\Lambda}$-dominated cosmology.

\section{Conclusions}

Equation~(\ref{eq:three}) implies that the change in the redshifts of
extragalactic objects due to the deceleration of the Universe is not far
out of reach of existing spectroscopic instrumentation.
For example, two sets of Keck HIRES observations of the \lya forest in the
spectrum of a hundred $B\sim 16$ mag quasars, separated by a decade, might
marginally allow for the statistical detection of the cosmic deceleration
signal.  The existence of some anomalously narrow line features in the \lya
forest can considerably improve the detection efficiency.

We find that peculiar accelerations due to large scale structure are
typically orders of magnitude smaller than the cosmic deceleration signal
($\sim H_0 c$).  Moreover, because of the random sign of their
contribution, local effects would also average out in a sufficiently large
sample of uncorrelated objects. In this sense, the cosmic deceleration
signal is similar to the microwave background anisotropies, who despite
their small amplitude (which posed a technological challenge for three
decades) appear to carry fundamental information about the Universe, and
not be confused by contamination.

The cosmic deceleration signal should be isotropic and have the particular
functional dependence on redshift predicted by equation~(\ref{eq:three}).
These characteristics can be used to identify its cosmic origin and to
infer the values of $H_0$, $\Omega_{\rm M}$, and $\Omega_\Lambda$ from it.
Even though detection of the cosmic deceleration signal appears mildly
impractical at present, further advances in spectroscopic techniques and
telescope size might allow it to be accessible in the future.  Such a
detection would complement the microwave anisotropy measurements, which
also probe the geometry of the Universe directly.  It is important to
perform both experiments independently as they cover different redshift
regimes, and could in principle even yield different results, due to the
existence of a time-dependent cosmological constant or decaying dark
matter.

\acknowledgements

I thank John Bahcall, Daniel Eisenstein, Bob Noyes, Ue-Li Pen, Bill Press,
George Rybicki, Ed Turner, and David Tytler for useful discussions, and
Dimitar Sasselov for suggesting the use of the Ly$\alpha$ forest.  This
work was supported in part by the NASA ATP grant NAG5-3085 and the Harvard
Milton fund.

\end{document}